\documentclass[prb,amsmath,groupedaddress,floatfix,twocolumn]{revtex4-1}

\usepackage{graphicx}
\usepackage{amssymb}
\usepackage{longtable}
\usepackage{rotating}
\usepackage{array}
\usepackage{multirow}
\usepackage{makecell}
\usepackage{booktabs}
\usepackage{color}
\usepackage{natbib}
\setcitestyle{numbers}
\setcitestyle{square}
\usepackage{hyperref}

\begin{document}

\title{Prediction and understanding of barocaloric effects in orientationally disordered materials from molecular dynamics simulations}

\author{Carlos Escorihuela-Sayalero,$^{1,2}$ Luis Carlos Pardo,$^{1,2}$ Michela Romanini,$^{1,2}$ Nicolas Obrecht,$^{3}$ Sophie Loehl\'{e},$^{3}$ Pol Lloveras,$^{1,2}$ Josep-Llu\'{i}s Tamarit,$^{1,2}$ Claudio Cazorla$^{1,2}$} 

\affiliation{$^{1}$Grup de Caracteritzaci\'{o} de Materials, Departament de F\'{i}sica, Universitat Polit\`{e}cnica de Catalunya, Campus Diagonal-Bes\`{o}s, Av. Eduard Maristany 10-14, 08019
Barcelona, Spain\\
$^{2}$Research Center in Multiscale Science and Engineering, Universitat Politècnica de Catalunya, Campus Diagonal-Bes\`{o}s, Av. Eduard Maristany 10-14, 08019
Barcelona, Spain\\
$^{3}$TotalEnergies OneTech, 69360 Salize, France
}

\begin{abstract}
Due to its high energy efficiency and environmental friendliness, solid-state cooling based on the barocaloric (BC) effect represents a promising alternative to traditional refrigeration technologies relying on greenhouse gases. Plastic crystals displaying orientational order-disorder solid-solid phase transitions have emerged among the most gifted materials on which to realize the full potential of BC solid-state cooling. However, a comprehensive understanding of the atomistic mechanisms on which order-disorder BC effects are sustained is still missing, and rigorous and systematic methods for quantitatively evaluating and anticipating them have not been yet established. Here, we present a computational approach for the assessment and prediction of BC effects in orientationally disordered materials that relies on atomistic molecular dynamics simulations and emulates quasi-direct calorimetric BC measurements. Remarkably, the proposed computational approach allows for a precise determination of the partial contributions to the total entropy stemming from the vibrational and molecular orientational degrees of freedom. Our BC simulation method is applied on the technologically relevant material CH$_{3}$NH$_{3}$PbI$_{3}$ (MAPI), finding giant BC isothermal entropy changes ($|\Delta S_{\rm BC}| \sim 10$~J~K$^{-1}$~kg$^{-1}$) under moderate pressure shifts of $\sim 0.1$~GPa. Intriguingly, our computational analysis of MAPI reveals that changes in the vibrational degrees of freedom of the molecular cations, not their reorientational motion, have a major influence on the entropy change that accompanies the order-disorder solid-solid phase transition.    
\end{abstract}
\maketitle

\section{Introduction}
\label{sec:intro}
Solid-state cooling represents an energy efficient and ecologically friendly solution to the environmental problems posed by conventional refrigeration technologies based on compression cycles of greenhouse gases \cite{manosa13,manosa2010,manosa17,cazorla19,lloveras21,Hou2022}. Upon small and moderate magnetic, electric and/or mechanical field shifts, promising caloric materials experience large adiabatic temperature variations ($|\Delta T| \sim 1$--$10$~K) as a result of phase transformations entailing large isothermal entropy changes ($|\Delta S| \sim 10$--$100$~J~K$^{-1}$~kg$^{-1}$). Solid-state cooling relies on such caloric effects to engineer multi-step refrigeration cycles. From an applied point of view, large $|\Delta T|$ and $|\Delta S|$ are both necessary for substantially and efficiently removing heat from a targeted system under recurrent switching on and off of a driving field. In terms of largest $|\Delta T|$ and $|\Delta S|$, mechanocaloric effects induced by uniaxial stress (elastocaloric effects) and hydrostatic pressure (barocaloric --BC-- effects) emerge as particularly encouraging \cite{manosa2010,manosa17,cazorla19,lloveras21}.

Colossal BC effects, $|\Delta S_{\rm BC}| \ge 100$~J~K$^{-1}$~kg$^{-1}$, and $|\Delta T_{\rm BC}| \ge 10$~K driven by pressure shifts of the order of $0.1$~GPa have been recently measured in several families of materials displaying orientational order-disorder solid-solid phase transitions  \cite{lloveras19,li19,aznar20,aznar21,Salvatori2022,li22,Sau2021,imamura20,lloveras21b,mason22}, thus achieving a major breakthrough in the field of solid-state cooling. Plastic crystals like neopentane derivatives \cite{lloveras19,li19,aznar20}, adamantane derivatives \cite{aznar21,Salvatori2022}, carboranes \cite{li22} and closo-borates \cite{Sau2021}, to cite some examples, conform the most representative family of disordered materials on which such a BC revolution has been realized. Colossal BC effects in plastic crystals have been intuitively rationalized in terms of large entropy changes predominantly originated by molecular orientational disorder stabilized under increasing temperature \cite{cazorla19c,hui22,li20,oliveira23} (as it is commonly assumed in molecular liquids \cite{luisca1,luisca2}). 

\begin{figure*}[ht!]
\includegraphics[width=0.95\linewidth]{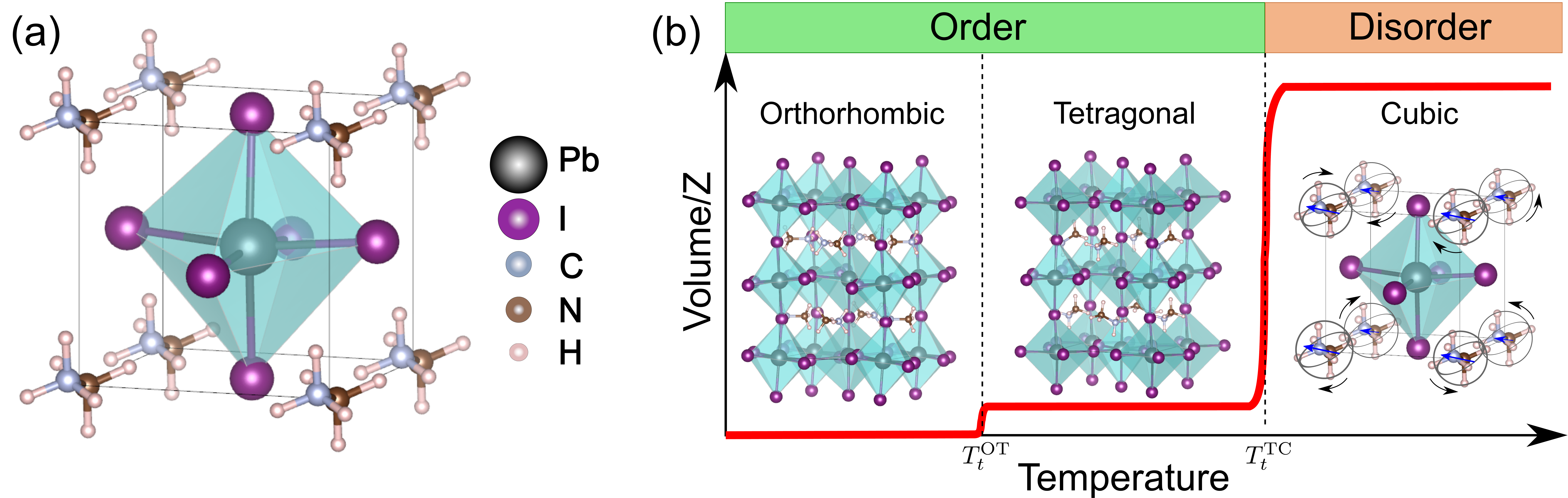}
\caption{\textbf{Characterization of CH$_{3}$NH$_{3}$PbI$_{3}$ (MAPI) and experimental sequence of $T$-induced phase transitions.} (a) Representation of the cubic unit cell of MAPI. (b) At $T_{t}^{OT} \approx 160$~K, MAPI undergoes a transformation from an orthorhombic to a tetragonal phase; at $T_{t}^{TC} \approx 330$~K, MAPI transforms into a cubic phase in which the molecular MA$^{+}$ ions are orientationally disordered \cite{cazorla22}. A non-negligible volume increase accompanies the order-disorder phase transition at $T_{t}^{TC}$.}
\label{fig1}
\end{figure*}

Nevertheless, despite these recent experimental advancements, a detailed atomistic understanding of colossal BC effects in plastic crystals is still missing and general theoretical approaches for assessing them have not been established. Owing to the first-order character of the involved order-disorder phase transition, a few authors have employed the Clausius-Clapeyron (CC) method in conjunction with atomistic molecular dynamics simulations to estimate $\Delta S_{\rm BC}$ \cite{Sau2021,hui22}. In the present context, however, the CC method presents several critical drawbacks: (1)~the quantity that is accessed is the phase transition entropy, $\Delta S_{t}$, which, although related, differs from the BC descriptor $\Delta S_{\rm BC}$, (2)~neither $\Delta T_{\rm BC}$ nor the temperature span over which BC effects are operative can be directly estimated, and (3)~partial entropy contributions stemming from different degrees of freedom (vibrational and orientational) cannot be determined. Alternatively, a few researchers have resorted to oversimplified \textit{ad hoc} $\Delta S_{\rm BC}$ analytical models with little to none predictive capability \cite{li20,oliveira23}. This lack of fundamental knowledge keeps hindering the rational design of disordered materials with improved BC performances, thus delaying the deployment of commercial solid-state cooling.

In this work, we advance towards the solution of the BC modelling conundrum in plastic crystals by developing and testing a facile computational approach based on molecular dynamics simulations. Our method emulates quasi-direct calorimetry measurements \cite{lloveras19,aznar20,aznar21,aznar17} and precisely provides the vibrational and molecular orientational contributions to the entropy without resorting to \textit{ad hoc} analytical models. As a case study, we apply our simulation BC approach to the technologically relevant material CH$_{3}$NH$_{3}$PbI$_{3}$ (MAPI, Fig.~\ref{fig1}a), a hybrid organic-inorganic perovskite that undergoes an order-disorder solid-solid phase transition under increasing temperature (Fig.~\ref{fig1}b) \cite{cazorla22}. In particular, we determined $\Delta S_{\rm BC}$ and $\Delta T_{\rm BC}$ under broad pressure and temperature conditions finding, for instance, a giant isothermal entropy change of $31$~J~K$^{-1}$~kg$^{-1}$ and an adiabatic temperature change of $9$~K for a pressure shift of $0.2$~GPa at temperatures below ambient. Intriguingly, our theoretical analysis concludes that the vibrational degrees of freedom of the molecular MA cations have a predominant role in the BC performance of MAPI, instead of the typically assumed molecular reorientations. This work establishes an insightful and predictive computational method for the estimation of BC effects in orientationally disordered systems like plastic crystals, hence it may be used to guide experiments and develop original solid-state refrigeration applications.

\section{Computational BC method}
\label{sec:methods}
\begin{figure}[ht!]
\includegraphics[width=0.95\linewidth]{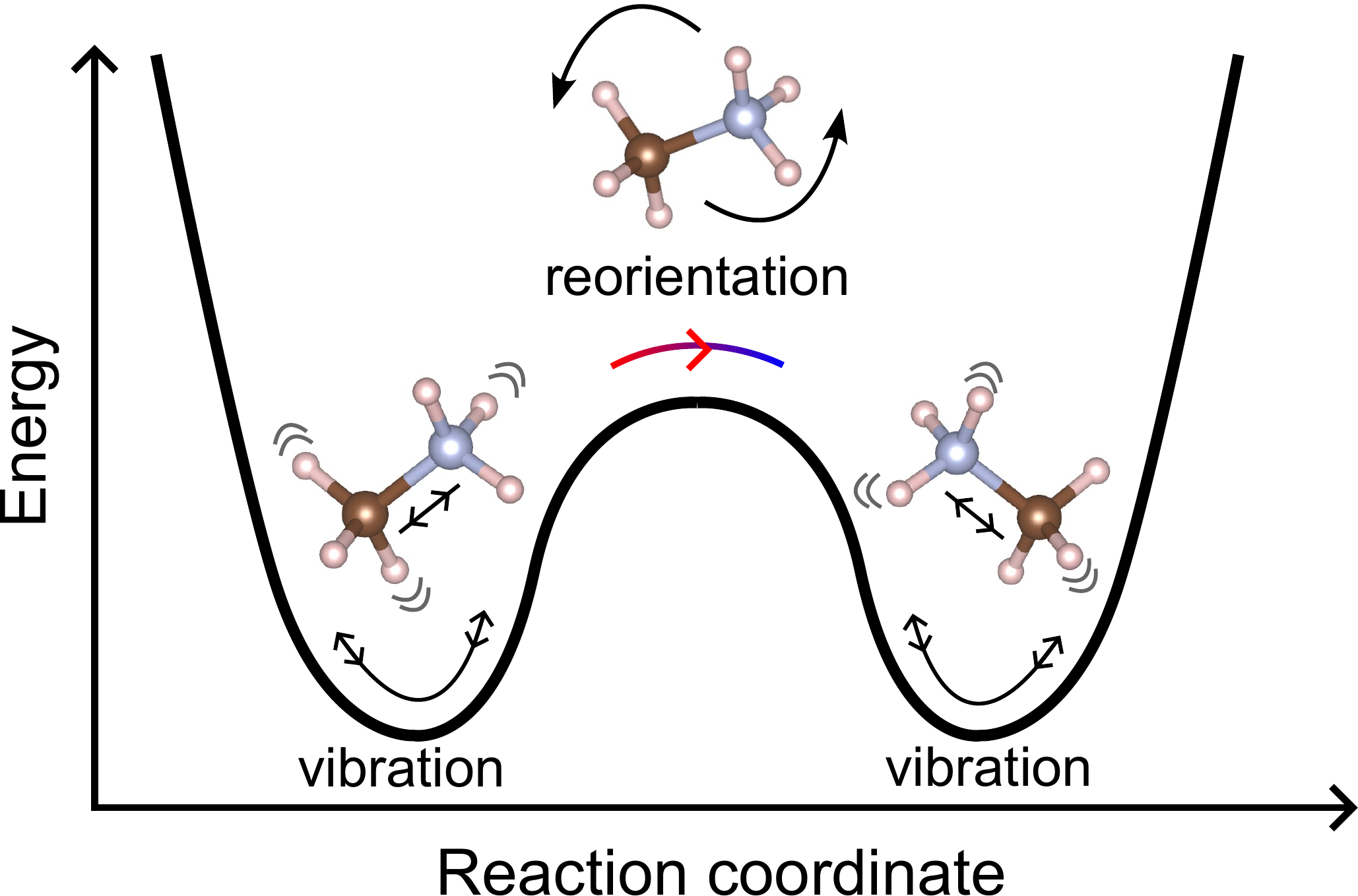}
\caption{\textbf{Sketch of $T$-induced molecular vibrational and orientational excitations in MAPI}. Vibrations entail periodic motion of the atoms around their equilibrium positions, corresponding to fluctuations around a particular local minimum in the potential energy surface. Orientational disorder, on the other hand, appears when thermal excitations are high enough for molecules to overcome the energy barriers between different local minima in the potential energy surface.}
\label{figsketch}
\end{figure}

In the present theoretical BC approach, the entropy of the low-$T$ ordered and high-$T$ orientationally disordered phases are determined as a function of pressure and temperature, $S(p,T)$, by assuming the relation:
\begin{equation}
    S(p,T) = S_{\rm vib}(p,T) + S_{\rm ori}(p,T)~,
    \label{eq:stot}
\end{equation}
where $S_{\rm vib}$ and $S_{\rm ori}$ are respectively the partial entropy contributions resulting from the vibrational and molecular orientational degrees of freedom (Fig.~\ref{figsketch}), and possible vibrational-orientational molecular couplings have been disregarded. In the low-$T$ ordered phase, $S_{\rm ori}$ is null while in the high-$T$ disordered phase is finite and positive. $S_{\rm vib}$, on the other hand, is finite and positive under any physically realizable thermodynamic conditions. Once $S(p,T)$ is determined for the low-$T$ and high-$T$ phases, the BC descriptors $\Delta S_{\rm BC}$ and $\Delta T_{\rm BC}$ are estimated just like it is done in quasi-direct calorimetric measurements \cite{lloveras19,aznar20,aznar21,aznar17} (Sec.~\ref{susbsec:quasi-direct}). Next, we explain in detail how to calculate each entropy term appearing in Eq.~(\ref{eq:stot}). 

\begin{figure*}[t]
\includegraphics[width=1.0\linewidth]{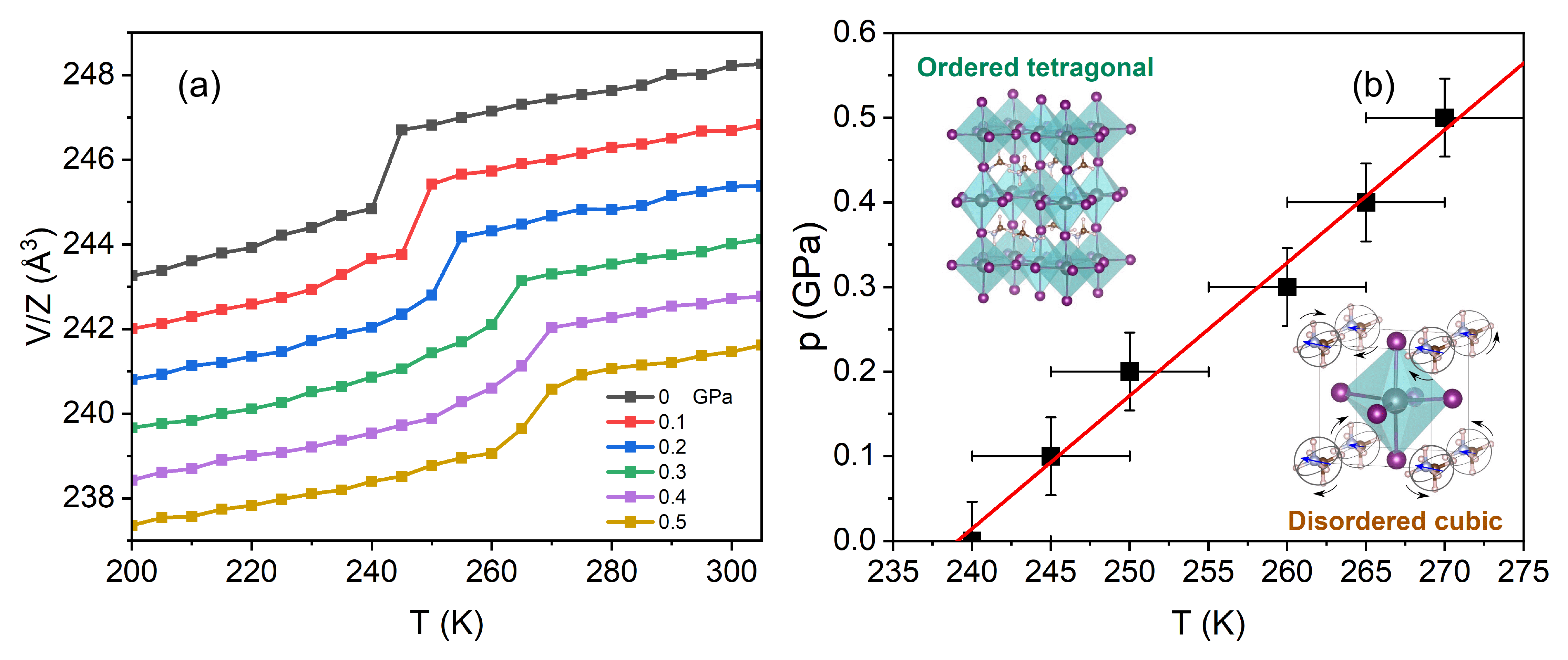}
\caption{\textbf{Order-disorder phase transition in MAPI characterized by $NpT$--MD simulations.} (a) Volume per formula unit expressed as a function of temperature at pressures $0.0 \le p \le 0.5$~GPa. The phase transition is characterized by a non-negligible change in volume. The lines are guides to the eye. (b) $p$--$T$ order-disorder phase boundary estimated for MAPI (red line, linear fit). The error bars result from the statistical fluctuations of the barostat and thermostat employed in the  $NpT$--MD simulations and the discreteness of the simulated $p$--$T$ conditions (Methods).}
\label{fig2}
\end{figure*}

\subsection{Vibrational entropy, $S_{\rm vib}$}
\label{subsec-svib}
The vibrational density of states, $\rho (\omega)$, besides providing information on the phonon spectrum of a crystal as a function of the vibrational frequency $\omega$, it also allows to estimate key thermodynamic quantities like the lattice free energy, $F_{\rm vib}$, and entropy, $S_{\rm vib}$. A possible way of calculating $\rho (\omega)$ from the outputs of $NpT$ molecular dynamics simulations ($NpT$--MD, Methods) consists in estimating the Fourier transform of the velocity autocorrelation function (VACF) \cite{Sagotra2019}.

Let $\mathbf{v_i}(t)$ be the velocity of the $i$-th particle expressed as a function of simulation time $t$. The VACF is given then by the expression:
\begin{equation}
    \text{VACF}(t) = \left\langle \mathbf{v}(0)\cdot \mathbf{v}(t) \right\rangle = \frac{1}{N} \sum_i^{N} \mathbf{v_i}(0)\cdot\mathbf{v_{i}}(t)~,
    \label{eq:vacf}
\end{equation}
where $\langle \cdots \rangle$ denotes statistical average in the $NpT$ ensemble and $N$ the number of particles in the system. Subsequently, the vibrational density of states can be estimated like \cite{Sagotra2019}: 
\begin{equation}
    \rho(\omega) = \int_0^{\infty}\text{VACF}(t)~e^{i\omega t}dt~.
    \label{eq:vdos}
\end{equation}
In the present case, we assume the vibrational density of states to fulfill the condition:
\begin{equation}
\int_0^{\infty} \rho(\omega)~d\omega = 3N~,
\label{eq:vdos-norm}
\end{equation}
where $3N$ corresponds to the number of lattice vibrations with real and positively defined phonon frequencies in the system. (In the case of assuming the normalization condition $\int_0^{\infty} \rho(\omega)~d\omega = 1$, an additional multiplicative factor $3N$ should be considered in the formulas appearing below in this subsection.) 

Upon determination of $\rho$, the vibrational free energy of the system can be straightforwardly estimated with the formula \cite{Cazorla2017}:
\begin{equation}
    F_{\rm vib} (p,T) = k_BT \int_0^{\infty} \ln{\left[ 2\sinh{\left(\frac{\hbar\omega}{2k_BT}\right)}\right]} \rho(\omega) d\omega~,
\label{eq:Fvib}
\end{equation}
where $k_{B}$ is the Boltzmann's constant. Consistently, the vibrational entropy, $S_{\rm vib}=-\frac{\partial F_{\rm vib}}{\partial T}$, adopts the expression:
\begin{eqnarray}
S_{\rm vib} (p,T) & = & -\int_0^{\infty} k_{B}\ln{\left[2\sinh{\left(\frac{\hbar\omega}{2k_BT}\right)}\right]} \rho(\omega) d\omega + \nonumber \\ 
& & \int_0^{\infty} \frac{\hbar\omega}{2T} \tanh^{-1}{\left(\frac{\hbar\omega}{2k_BT}\right)} \rho(\omega) d\omega~.
\label{eq:svib}
\end{eqnarray}
It is worth noting that the dependence on pressure and temperature of the thermodynamic quantities in Eqs.~(\ref{eq:Fvib})--(\ref{eq:svib}) are implicitly contained in $\rho (\omega)$, since this function is directly obtained from the outputs of atomistic $NpT$--MD simulations.

\subsection{Molecular orientational entropy, $S_{\rm ori}$}
\label{subsec-sori}
For a continuous random variable $x$ with probability density $f(x)$, its entropy is defined as \cite{informationtheory}:
\begin{equation}
S = - \int_{X} f(x) \log{f(x)}~dx~,
\label{eq:shanon}
\end{equation}
where the integral runs over all possible values of $x$. If $x$ represents a microstate characterizing a particular thermodynamic macrostate, then the following Gibbs entropy can be defined for the system of interest \cite{pathria1972}:  
\begin{equation}
S_{G} = -k_{B} \int_{X} f(x) \log{f(x)}~dx~.
\label{gibbs-entropy}
\end{equation}
 
In a orientationally disordered crystal, molecules reorient in a quasi-random fashion as shown by the time evolution of their polar ($\theta$) and azimuthal ($\varphi$) angles as referred to a fixed arbitrary molecular axis (e.g., the C--N bond in the methylammonium molecule at a given time). By assuming the molecules in the crystal to be independent one from the other, one may then estimate a probability density function for their orientation, $f(\theta, \varphi)$, by considering, for instance, the atomistic trajectories generated during long $NpT$--MD simulations. In such a case, one can define the following three-dimensional molecular angular entropy (in which the molecules length are considered fixed):
\begin{equation}
        S_{\rm ang} = -k_{B} \int_{\Theta, \Phi} f(\theta, \varphi)\log{f(\theta, \varphi)}~d\cos\theta d\varphi~.
\label{eq:sang-cont}
\end{equation}

In practice, the calculation of $S_{\rm ang}$ entails the construction of histograms for which the continuous polar variables are discretised, $(\theta, \varphi) \to (\theta_{i}, \varphi_{j})$, and  bin probabilities are defined like:
\begin{equation}
        p_{i}(\theta_{i}, \varphi_j) \approx f(\theta_{i}, \varphi_j) \Delta{\cos\theta} \Delta\varphi~,
\label{eq:binprob}       
\end{equation}
where $\Delta{\cos\theta} \Delta\varphi$ corresponds to the area of an histogram bin (assumed to be constant here). Consistently, one can then rewrite $S_{\rm ang}$ in the discretised form \cite{informationtheory}:
\begin{equation}
        S_{\rm ang} = -k_{B} \sum_{i} p_{i} \log{p_{i}} + k_{B} \log{\left(\Delta{\cos\theta} \Delta\varphi  \right)}~.
\label{eq:sang-disc}        
\end{equation}

The angular entropy defined in the equation above, however, is not necessarily equal to the entropy associated with molecular orientational disorder alone since other molecular degrees of freedom (e.g., molecular fluctuations and librations, which are of vibrational nature) are also inevitably captured by the angular probability density function $f(\theta, \varphi)$ deduced from $NpT$--MD simulations (Sec.~\ref{susbsec:quasi-direct}). A practical way of getting rid of most spurious contributions to the molecular orientational entropy in the high-$T$ disordered phase consists in offsetting $S_{\rm ang}$ by the maximum value attained in the low-$T$ ordered phase, in which molecular reorientations do not actually occur at an appreciable rate. Due to the fact that the largest molecular oscillations in the low-$T$ ordered phase occur close to the order-disorder phase transition temperature, $T_{t}$, an appropriate molecular orientational entropy then can be defined like:    
\begin{equation}
    S_{\rm ori}(p,T) =
    \left\{
    \begin{array}{lr}
    S_{\rm ang} (p,T) - S_{\rm ang} (p,T_t) & T \ge T_{t} \\
    0 & T < T_{t}
    \end{array}
    \right.,
\label{eq:S_ori}
\end{equation}
which is the expression that we have used throughout this work.

\begin{table*}[t]
\begin{tabular}{cccccc}
\hline
\hline
\quad $p$ \quad & \quad $T_{t}$ \quad  & \quad $dT_{t}/dp$ \quad & \quad $\Delta V_{t}/Z$ \quad & \quad $\Delta V_{t}/V$ \quad & \quad $\Delta S_{t}$ \quad \\
(GPa) & (K) & (K~GPa$^{-1}$) & (\AA$^{3}$) & (\%)& (J~K$^{-1}$~kg$^{-1}$)  \\
\hline
0.0 & 240 & 63.7 & 1.86 & 0.8 & 28.4 \\
0.1 & 245 & 63.7 & 1.66 & 0.7 & 25.4 \\
0.2 & 250 & 63.7 & 1.37 & 0.6 & 19.2 \\
0.3 & 260 & 63.7 & 1.04 & 0.3 & 15.9 \\
0.4 & 265 & 63.7 & 0.90 & 0.4 & 13.8 \\
0.5 & 270 & 63.7 & 0.94 & 0.4 & 14.3 \\
\hline
\hline
\end{tabular}
\caption{\textbf{Order-disorder solid-solid phase transition in MAPI characterized by $NpT$--MD simulations.} Results were obtained with the MAPI force field developed by Mattoni and co-workers \cite{mattoni_model1,mattoni_model2,mattoni_model3}. The phase-transition entropy changes, $\Delta S_{t}$, were estimated with the Clausius-Clapeyron relation. The numerical uncertainties of the reported pressures and transition temperatures amount to $0.05$~GPa and $5$~K, respectively.}
\label{table1}
\end{table*}

\section{Results}
\label{sec:results}
We used the method introduced in Sec.~\ref{sec:methods} to estimate BC effects in bulk CH$_{3}$NH$_{3}$PbI$_{3}$ (MAPI, Fig.~\ref{fig1}), a highly promising photovoltaic material with exceptional and highly tunable optoelectronic properties \cite{mapi1,mapi2,mapi3}. The reasons for this material selection are several. First, at a temperature of $\approx 330$~K, MAPI undergoes a structural transformation from a tetragonal to a cubic phase in which the methylammonium molecular cations (MA$^{+}$) are orientationally disordered \cite{cazorla22} (Fig.~\ref{fig1}b). MAPI, therefore, conforms to the broad definition of orientationally disordered crystal despite being generally, and more conveniently, classified as a hybrid organic-inorganic perovskite. Second, previous experimental and theoretical works have already addressed the existence of possible caloric effects in MAPI \cite{caloric-mapi1,caloric-mapi2}, thus our results may be compared with those of other studies. And third, a reliable force field has been already reported and tested for MAPI in the literature \cite{mattoni_model1,mattoni_model2,mattoni_model3} that allows to perform thousands-of-atoms $NpT$-MD simulations in a very efficient manner (Methods).   
The organization of this section is as follows. First, we characterize the molecular order-disorder (OD) phase transition in MAPI under broad temperature and pressure conditions by means of atomistic $NpT$--MD simulations performed with the force field developed by Mattoni and co-workers \cite{mattoni_model1,mattoni_model2,mattoni_model3}. Next, we analyse the entropy change associated with the phase transition of interest by using the well-known Clausius-Clapeyron equation \cite{Sau2021}. Finally, we explain in detail the estimation of the partial entropy contributions $S_{\rm vib} (p,T)$ and $S_{\rm ori} (p,T)$ and of the barocaloric descriptors $\Delta S_{\rm BC}$ and $\Delta T_{\rm BC}$.

\subsection{Order-disorder phase transition in MAPI}
\label{subsec:pd}
Figure~\ref{fig2} shows the theoretical characterization of the OD phase transition in MAPI at pressures of $0 \le p \le 0.5$~GPa as obtained from $NpT$-MD simulations (Methods). In the high-$T$ phase the molecular MA cations are orientationally disordered while in the low-$T$ phase they arrange orderly in an antiferroelectric pattern \cite{antiferro-mapi}. At zero pressure, this phase transition experimentally occurs near room temperature and is accompanied by a relative volume expansion of $0.2$--$0.3$\% \cite{mapi-exp1,mapi-exp2}. The MAPI force field employed in this work \cite{mattoni_model1,mattoni_model2,mattoni_model3} qualitatively mimics such a OD phase transition, however, it does not quantitatively reproduce the experimental $T_{t}$ and transition volume expansion value. 

Figure~\ref{fig2}a shows the evolution of the volume per formula unit calculated across the OD MAPI phase transformation at different pressure conditions. In the $NpT$-MD simulations, the transition temperature is identified by a sudden increase in volume that coincides with the stabilization of molecular orientational disorder (confirmed by rapidly decaying angular autocorrelation functions estimated under the same conditions \cite{Sau2021}, Supplementary Fig.~1). In our zero-pressure simulations, we calculated a relative volume expansion of $0.8$\% at a transition temperature of $\approx 240$~K, values that are appreciably larger and smaller, respectively, than the corresponding experimental figures. Nevertheless, since the main objective of the present study is to develop and test a solid BC computational approach, such a lack of quantitative force field accuracy should not be considered as a limiting factor (Sec.~\ref{susbsec:quasi-direct}).   

Figure~\ref{fig2}b shows the dependence of $T_{t}$ estimated for MAPI under pressure. When considering the numerical uncertainties in our $NpT$-MD simulations, a linear $p$-dependence of the OD transition temperature clearly emerges. In particular, a first-order polynomial fit of the form $T_{t} (p) = 239.2 + 63.7p$, in which the temperature and pressure are respectively expressed in units of K and GPa, is found to best reproduce the computed transition temperatures. The barocaloric coefficient $dT_{t}/dp$ deduced from the simulations roughly amounts to $60$~K~GPa$^{-1}$, which in this case is in fairly good agreement with the experimental values of $76$--$35$~K~GPa$^{-1}$ reported at zero-pressure conditions \cite{caloric-mapi1}. Table~\ref{table1} summarizes the key features of the pressure-induced OD MAPI phase transition as obtained from our $NpT$-MD simulations, including the phase transition entropy change, $\Delta S_{t}$.

\begin{figure*}[t]
\includegraphics[width=1.0\textwidth]{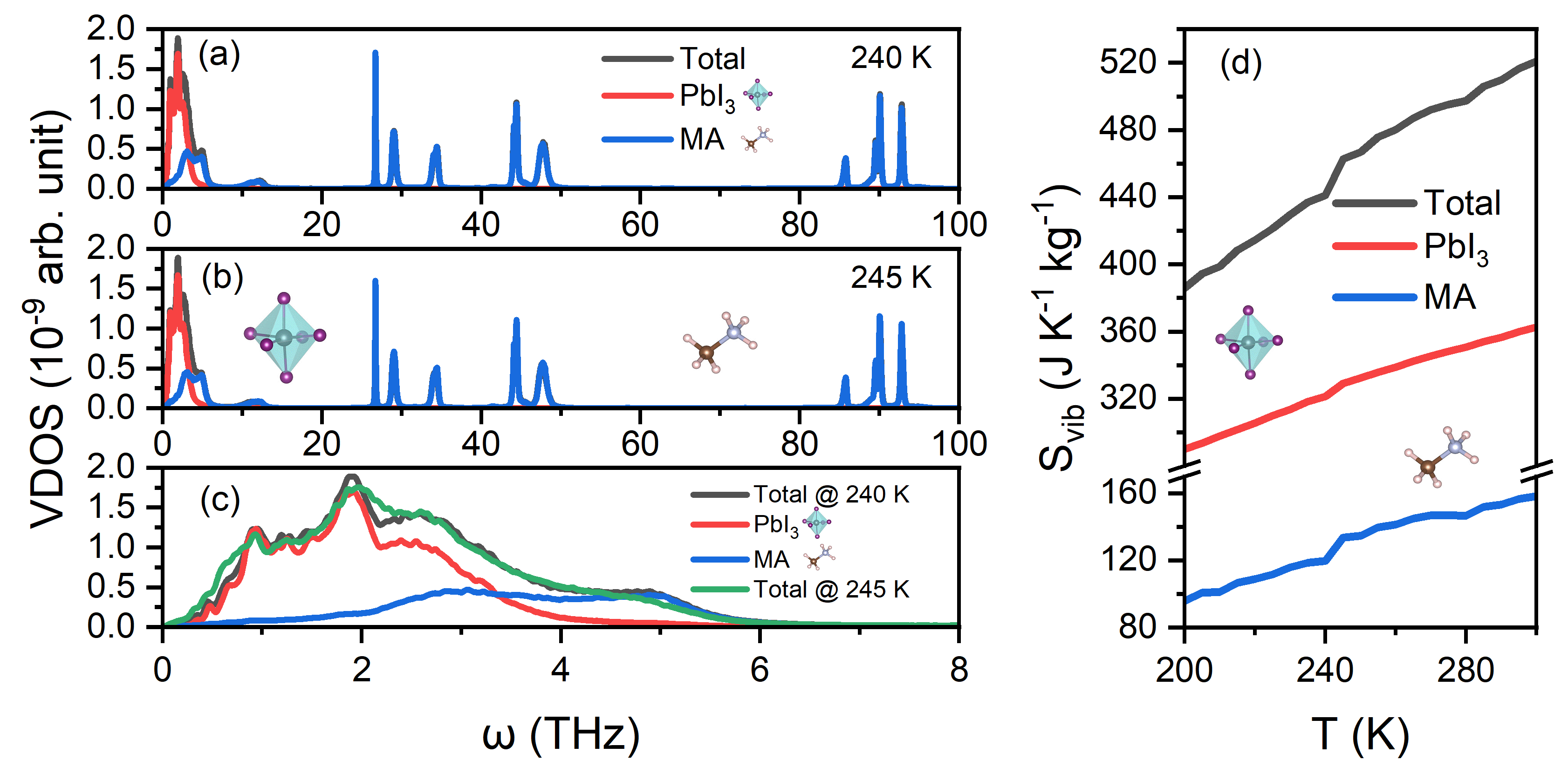}
\caption{\textbf{Vibrational density of states and vibrational entropy, $S_{\rm vib}$, estimated for MAPI.} Total VDOS and partial contributions stemming from the PbI$_3$ and MA motifs at zero pressure and $T = 240$~K (a) and $245$~K (b). (c) Comparison of the VDOS obtained for low vibrational frequencies at temperatures right below and above the MAPI order-disorder phase transition. (d) Vibrational entropy calculated for MAPI as a function of temperature at zero pressure; partial contributions from the PbI$_3$ and the MA groups are also shown.}
\label{fig3}
\end{figure*}

\subsection{The Clausius-Clapeyron method}
\label{subsec:cc}
The well-known Clausius-Clapeyron equation,
\begin{equation}
        \Delta S_{t} = \left(\frac{dT_t}{dp}\right)^{-1} \Delta V_{t}~,
\label{eq:cc-formula}
\end{equation}
allows for the estimation of the entropy change accompanying a phase transition, $\Delta S_{t}$, based on the knowledge of the slope of its $p$--$T$ phase boundary and the concomitant change in volume, $\Delta V_{t}$. By applying this relation to the results of our $NpT$-MD simulations, we obtained the phase transition entropy changes summarized in Table~\ref{table1}. The predicted entropy change approximately amounts to $28$~J~K$^{-1}$~kg$^{-1}$ at zero pressure and steadily decreases under increasing compression (e.g., $\Delta S_{t} = 19.2$~J~K$^{-1}$~kg$^{-1}$ at $0.2$~GPa). Such a $p$-induced phase transition entropy decrease is due exclusively to the $\Delta V_{t}$ reduction originated by pressure (Table~\ref{table1}) since, to a good approximation, the slope of the $p$--$T$ phase boundary in our $NpT$-MD simulations is constant (Fig.~\ref{fig2}b). 

In the absence of external applied pressure, the experimental $\Delta S_{t}$ value reported for MAPI amounts to $15.65$~J~K$^{-1}$~kg$^{-1}$ \cite{caloric-mapi1}, which is roughly two times smaller than the one estimated here with $NpT$-MD simulations (Table~\ref{table1}). This noticeable difference follows from the $dp/dT_t$ and $\Delta V_{t}$ discrepancies found between the experiments and our calculations, which have been explained in the previous paragraphs. 

As it can be appreciated in Fig.~\ref{fig2}b, the assessed $T_{t}$'s come with some numerical uncertainties (i.e., error bars in the figure) that result from the fluctuations introduced by the thermostat and barostat employed in the $NpT$-MD simulations along with the discreteness of the simulated $p$--$T$ conditions (Methods). The presence of pre-transitional effects also manifests at the highest simulated pressures, as shown by the $T$-dependence of the volume system obtained across the phase transition at $p = 0.4$ and $0.5$~GPa (i.e., the volume variations become less abrupt making more difficult the identification of $\Delta V_{t}$, Fig.~\ref{fig2}a). These errors, which to a certain extent are unavoidable in $NpT$-MD simulations, lead to inaccuracies in the estimation of $dp/dT_t$ and phase transition volume changes, and inevitably propagate to $\Delta S_{t}$. In the following section, we present a more refined method for the calculation of phase transition entropy changes, and entropy curves in general, based on the computational strategy introduced in Sec.~\ref{sec:methods}, thus overcoming the usual numerical limitations of the Clausius-Clapeyron approach \cite{Sau2021,hui22}.

\begin{figure*}[t]
\includegraphics[width=1.0\linewidth]{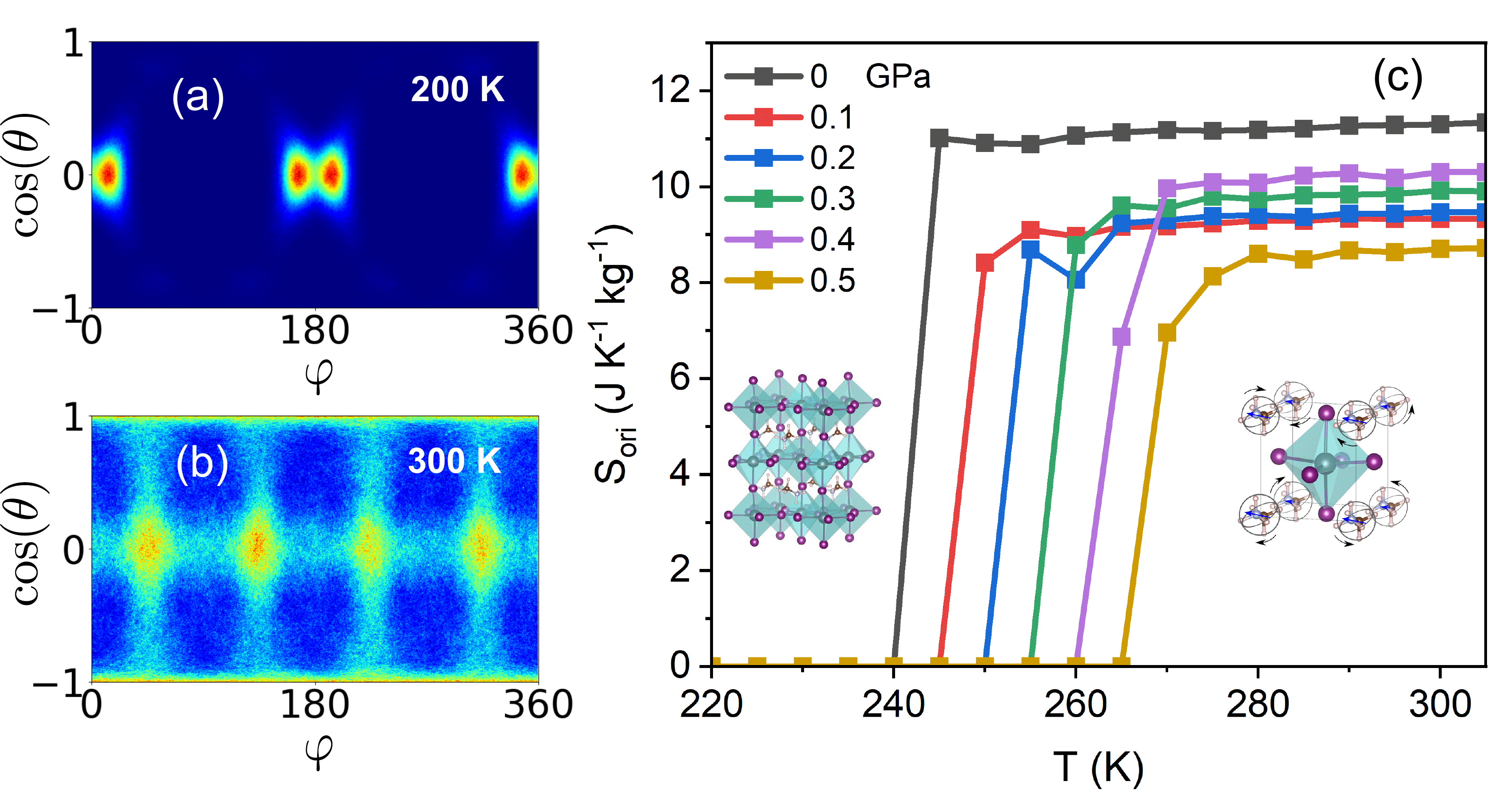}
\caption{\textbf{Characterization of the angular distribution of molecular cations in MAPI and resulting orientational entropy.} (a)~Angular probability density for the molecular MA cations in the ordered phase at $T = 200$~K. High-probability regions are represented with red color whereas low-probability regions with dark blue color. (b)~Angular probability density for the molecular MA cations in the disordered phase at $T = 300$~K. (c)~Molecular orientational entropy, $S_{\rm ori}$, calculated at different temperatures and pressures. The lines are guides to the eye.}
\label{fig4}
\end{figure*}

\subsection{Estimation of BC effects in MAPI}
\label{susbsec:quasi-direct}
\subsubsection{Vibrational entropy}
\label{subsec:svib}
Figure~\ref{fig3} shows the vibrational density of states (VDOS) and partial organic and inorganic contributions estimated for MAPI at zero pressure and temperatures slightly above and below the simulated OD transition point (i.e., $245$ and $240$~K). The VDOS contribution assigned to the inorganic part, namely, the PbI$_{3}$ octahedra, is clearly dominant in the low-frequency range $0 \le \nu \le 5$~THz (Figs.~\ref{fig3}a--c). This result follows from the fact that the lighter atoms, which typically vibrate at higher frequencies, entirely reside in the organic molecules. Consistently, the range of moderate and high frequencies, $\nu \ge 5$~THz, is mostly governed by the MA cation vibrations.

Albeit the VDOS enclosed in Figs.~\ref{fig3}a,b may seem quite similar at first glance, there are significant differences among them. In particular, the high-$T$ disordered phase accumulates more phonon modes in the low-frequency range $0 \le \nu \le 2$~THz than the low-$T$ ordered phase (Fig.~\ref{fig3}c and Supplementary Fig.~2). This effect has a strong impact on the vibrational entropy of the system, $S_{\rm vib}$, which is significantly larger in the high-$T$ disordered phase. 

Figure~\ref{fig3}d shows the $S_{\rm vib}$ estimated at zero pressure as a function of temperature. A clear surge in the vibrational entropy is observed at the OD transition point, which amounts to $\Delta S_{\rm vib} = 21.66$~J~K$^{-1}$~kg$^{-1}$. The primary contribution to such a vibrational entropy increase comes from the molecular cations, which is equal to $13.93$~J~K$^{-1}$~kg$^{-1}$ and almost twice as large as that calculated for the inorganic part ($7.73$~J~K$^{-1}$~kg$^{-1}$). Thus, although the low-frequency range in the VDOS is dominated by the inorganic anions, the organic MA cations have a larger influence on the vibrational entropy change associated with the order-disorder phase transition, $\Delta S_{\rm vib}$. 

\begin{figure*}[t]
\includegraphics[width=1.0\linewidth]{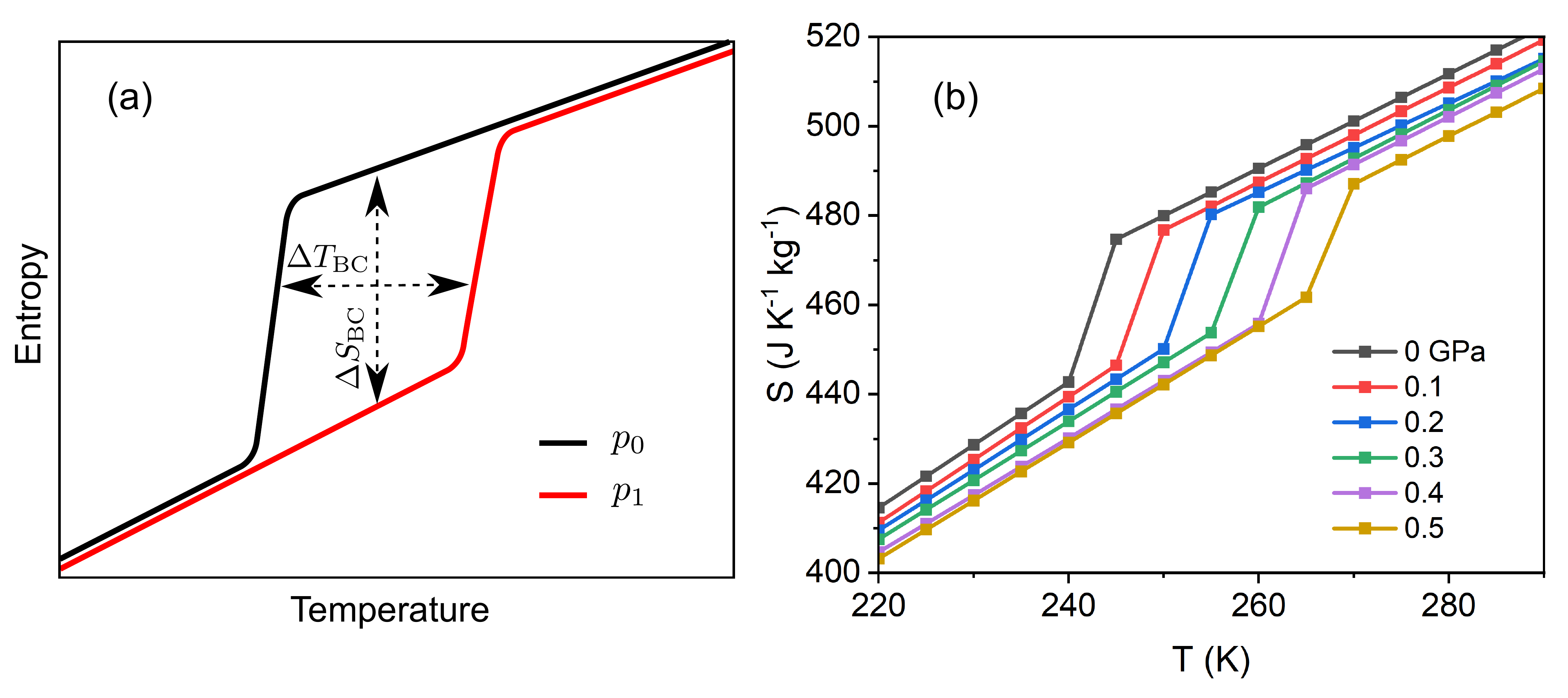}
\caption{\textbf{Quasi-direct estimation of BC descriptors and total entropy curves calculated for MAPI.} (a)~The barocaloric descriptors $\Delta S_{\rm BC}$ and $\Delta T_{\rm BC}$ can be directly assessed from the $S$--$T$ curves obtained at different pressures, as it is schematically shown in the figure. (b)~Entropy curves calculated for MAPI from $NpT$--MD simulations as a function of temperature and considering different fixed pressures.}
\label{fig5}
\end{figure*}

\subsubsection{Molecular orientational entropy}
\label{subsec:srot}
Figures~\ref{fig4}a,b show the angular probability density associated with the orientation of a molecular MA cation in MAPI at temperatures below and above the simulated OD transition point (i.e., averaged over all the molecules in the simulation cell). The coordinates $(\varphi, \theta)$ represent the azimuthal and polar angles for a MA molecule considering its C--N bond axis and the crystallographic axes as the rest reference system. In the low-$T$ ordered phase, the molecules do not reorient but tightly fluctuate around their preferential orientations (represented by sharp red spots in Fig.~\ref{fig4}a), which alternate from cation to cation forming a global antiferroelectric pattern \cite{antiferro-mapi}. In the high-$T$ disordered phase, on the contrary, elongated regions of roughly equal probability appear that connect the preferential molecular orientations (i.e., the blurry greenish areas in Fig.~\ref{fig4}b). These equiprobable density regions represent the actual paths through which the MA cations reorient (a three-dimensional sketch of those rotational tracks is provided in the Supplementary Fig.~3). 

\begin{figure*}[t]
\includegraphics[width=1.0\linewidth]{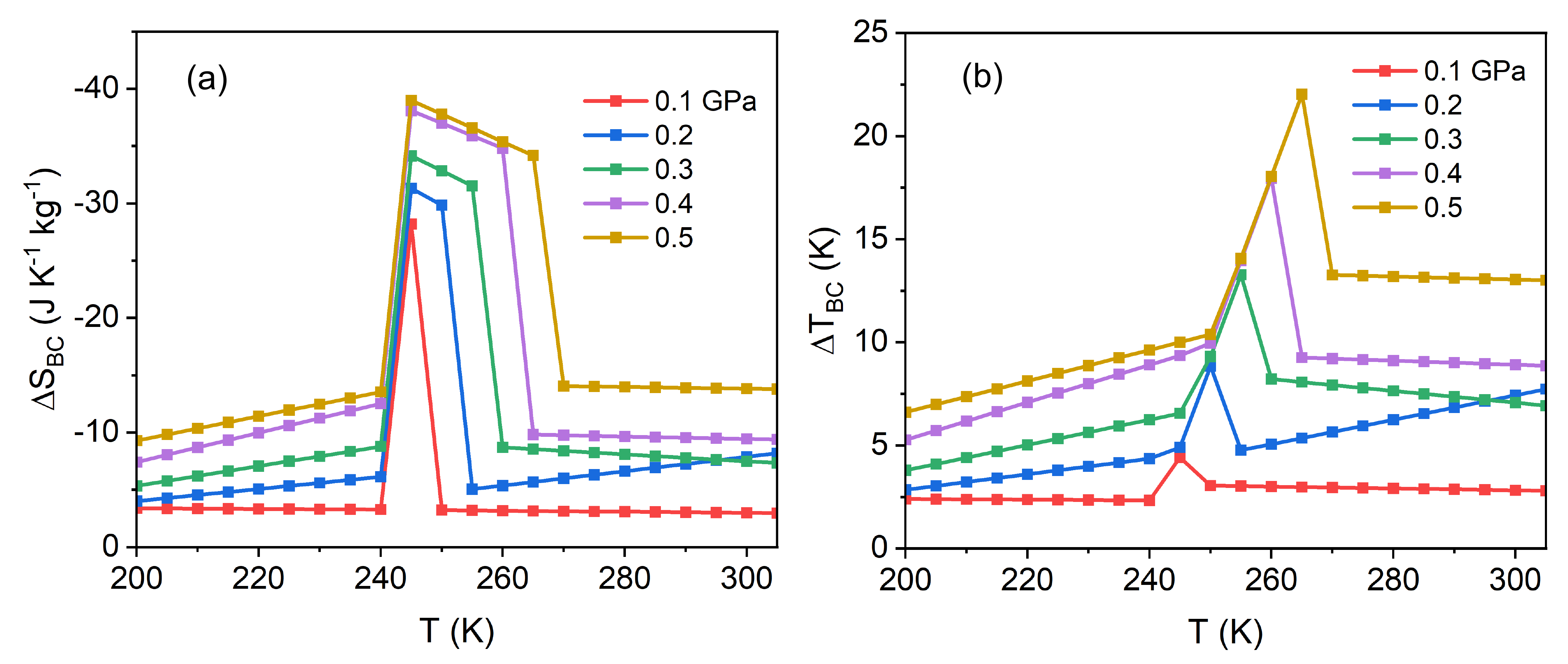}
\caption{\textbf{Theoretical estimation of barocaloric effects in MAPI.} (a)~Isothermal entropy change, $\Delta S_{\rm BC}$, calculated as a function of $T$ and $\Delta p$. (b)~Adiabatic temperature change, $\Delta T_{\rm BC}$, calculated as a function of $T$ and $\Delta p$. BC effects at conditions different from the transition points are not null essentially due to the thermal expansion of the material \cite{aznar17}.}
\label{fig6}
\end{figure*}

It is worth noting that in the high-$T$ disordered phase the fluctuations of the MA$^{+}$ cations around their equilibrium orientations are very broad, as it is deduced from the enlarged areas of maximum probability in Fig.~\ref{fig4}b (in comparison to those shown in Fig.~\ref{fig4}a). These fluctuations conform to molecular librations with very low vibrational frequencies due to the large effective mass associated with them. Such a collective librational dynamics appears to be the main responsible for the vibrational entropy gain identified for the high-$T$ disordered phase in the previous Sec.~\ref{subsec:svib}.   

Figure~\ref{fig4}c shows the orientational entropy of the MA$^{+}$ cations, $S_{\rm ori}$, estimated as a function of temperature and pressure by employing the method introduced in Sec.~\ref{subsec-sori}. At temperatures below the theoretical transition point, $S_{\rm ori}$ is null since molecular reorientations do not occur at an appreciable rate. At temperatures above $T_{t}$, on the other hand, $S_{\rm ori}$ is sizeable and practically independent of temperature. The orientational entropy plateau that is rapidly attained at $T > T_{t}$ can be understood in terms of two different facts: (1)~the paths through which the molecules reorient in the disordered phase are not significantly affected by increasing thermal excitations, and (2)~although molecular reorientations occur at a higher rate at higher temperatures, the probability of finding a molecule around an equilibrium orientation or in a transient orientational state is pretty independent of temperature. It is worth noting that the behaviour described here for $S_{\rm ori}$ is plainly different from that of a rigid molecular free rotor \cite{li20,oliveira23}, for which the entropy monotonically increases under increasing temperature (Discussion).

The pressure dependence of the molecular orientational entropy also can be appreciated in Fig.~\ref{fig4}c. In the high-$T$ ordered phase, compression only moderately reduces $S_{\rm ori}$. For instance, at zero pressure it approximately amounts to $11$~J~K$^{-1}$~kg$^{-1}$ while at $p = 0.5$~GPa is equal to $\approx 8.5$~J~K$^{-1}$~kg$^{-1}$. In the compression interval $0.1 \le p \le 0.4$~GPa, however, $S_{\rm ori}$ behaves quite irregularly and adopts values of $9$--$10$~J~K$^{-1}$~kg$^{-1}$. Such fluctuations are due to the numerical uncertainties in our calculations and the mild influence of pressure on $S_{\rm ori}$.  

We performed a conclusive test to assess the accuracy of our $S_{\rm ori}$ calculation method. At zero pressure, the entropy change associated with the order-disorder phase transition can be directly obtained from the $NpT$-MD simulations since the Gibbs free energy equality between the two involved phases, $G^{O} (0,T_{t}) = G^{D} (0, T_{t})$, leads to the relation $\Delta S_{t} = \Delta U_{t} / T_{t}$, where $\Delta U_{t} \equiv U^{D} - U^{O}$ is the internal energy difference between the disordered and ordered phases at the specified $p$--$T$ conditions. Then, by considering that $\Delta S_{t} = \Delta S_{\rm vib} + \Delta S_{\rm ori}$ and assuming that $\Delta S_{\rm vib}$ is perfectly evaluated with the method described in Sec.~\ref{subsec-svib}, it is possible to directly compute $S_{\rm ori}$ without the need of performing integrations over polar probability maps. By proceeding so, we obtained a minute discrepancy of only $0.5$~J~K$^{-1}$~kg$^{-1}$ with respect to the $S_{\rm ori}$ value obtained with the method introduced in Sec.~\ref{subsec-sori} and shown in Fig.~\ref{fig4}c. Similar excellent agreement between the two independent $S_{\rm ori}$ evaluation approaches was also found for other transition points obtained under $p \neq 0$ conditions (in this latter case, $\Delta S_{t} = 1/T_{t} \left[\Delta U_{t} + p\Delta V_{t} \right]$). Therefore, based on these findings we may conclude that (1)~for any arbitrary $p$--$T$ conditions, not necessarily ascribed to a phase transition point, our $S_{\rm ori}$ estimation method based on $NpT$-MD simulations is fully reliable, and (2)~the neglect of possible vibrational-orientational molecular couplings in Eq.~(\ref{eq:stot}) appears to be reasonable (at least, for MAPI).

\subsubsection{Total entropy curves and estimation of BC effects}
\label{subsec:deltast}
We calculated the total entropy of MAPI as a function of temperature and pressure by adding up the vibrational and orientational contributions reported in the previous two sections, namely, $S(p,T) = S_{\rm vib} (p,T) + S_{\rm ori} (p,T)$. The resulting $S(p,T)$ curves are very well behaved, as it is explicitly shown in Fig.~\ref{fig5}b. In analogy to quasi-direct calorimetry experiments \cite{lloveras19,aznar20,aznar21,aznar17}, the estimation of the barocaloric isothermal entropy change, $\Delta S_{\rm BC}$, and adiabatic temperature change, $\Delta T_{\rm BC}$, now turns out to be quite straightforward, as it is schematically shown in Fig.~\ref{fig5}a. In particular, we have that:
\begin{eqnarray}
\Delta S_{\rm BC}(T, 0 \rightarrow p) & = & S(T, p) - S(T, 0) \\
\Delta T_{\rm BC}(T, 0\rightarrow p)  & = & T_{\rm f}(S, p) - T(S, 0)~,
\end{eqnarray}
where $T_{\rm f}$ stands for the temperature fulfilling the condition $S(T_{\rm f}, p) = S(T, 0)$ (Fig.~\ref{fig5}a).

Figure~\ref{fig6} shows our barocaloric $\Delta S_{\rm BC}$ and $\Delta T_{\rm BC}$ results obtained in MAPI for pressure shifts of $0 \le \Delta p \le 0.5$~GPa. In Fig.~\ref{fig6}a, it is clearly appreciated that under increasing compression the maximum $\Delta S_{\rm BC}$ value also increases. For instance, the maximum isothermal entropy change obtained for $\Delta p = 0.1$~GPa amounts to $-28.19$~J~K$^{-1}$~kg$^{-1}$ whereas for $\Delta p = 0.5$~GPa is equal to $-38.99$~J~K$^{-1}$~kg$^{-1}$. It is worth noting that for $\Delta p \ge 0.4$~GPa the growth rate of $\Delta S_{\rm BC}$ is found to diminish drastically. The temperature span over which BC effects remain sizeable also increases under increasing pressure shifts, changing for instance from $\approx 10$~K for $\Delta p = 0.2$~GPa to $\approx 20$~K for $0.5$~GPa. These isothermal entropy shift values, although probably are somewhat overestimated due to the limitations of the employed force field evidenced in Secs.~\ref{subsec:pd}--\ref{subsec:cc}, can be regarded as giant since they are of the order of $10$~J~K$^{-1}$~kg$^{-1}$ \cite{lloveras21}. 

The evolution of $\Delta T_{\rm BC}$ under increasing $\Delta p$ is reported in Fig.~\ref{fig6}b. As already expected from the previous isothermal entropy results, the maximum adiabatic temperature change is also significantly enhanced under increasing compression. In particular, the $\Delta T_{\rm BC}$ peak changes from $4.41$~K for a pressure shift of $\Delta p = 0.1$~GPa to $22.03$~K for $0.5$~GPa. In this case, the size of $\Delta T_{\rm BC}$ is not found to saturate at the highest considered pressure. It is worth noting here that at conditions other than the phase transition points the estimated BC effects are not null (Fig.~\ref{fig6}). These results follow from the thermal expansion properties of MAPI in the ordered and disordered phases \cite{aznar17}, not from the OD phase transition itself.   

\begin{figure}[t]
\includegraphics[width=1.0\linewidth]{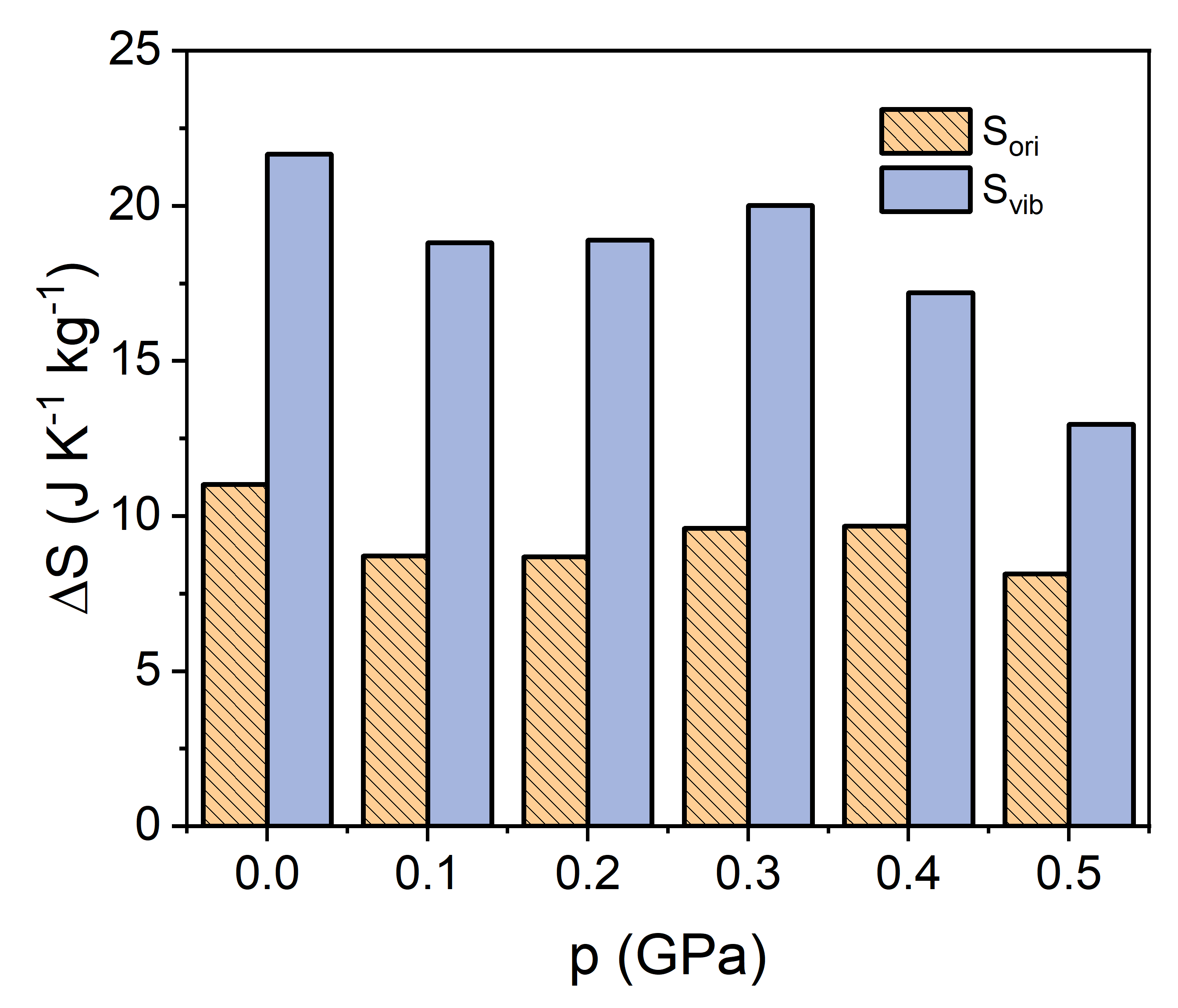}
\caption{\textbf{Contributions to the total entropy change associated with the order-disorder phase transition in MAPI.} $\Delta S_{\rm vib}$ and $\Delta S_{\rm ori}$ represent the total vibrational and molecular orientational entropies, respectively, and it follows that $\Delta S_{t} = \Delta S_{\rm vib} + \Delta S_{\rm ori}$.}
\label{fig7}
\end{figure}

In a previous work, Liu and Cohen investigated the possible existence of elastocaloric effects in MAPI (i.e., induced by uniaxial stress, $\sigma$) based on similar molecular dynamics simulations than reported here \cite{caloric-mapi2} (e.g., employed also the force field developed by Mattoni and co-workers \cite{mattoni_model1,mattoni_model2,mattoni_model3}). Those authors used a direct elastocaloric estimation approach that requires of very long simulation times to attain adiabatic conditions and which cannot straightforwardly provide elastocaloric $\Delta S$ results. Interestingly, Liu and Cohen reported an elastocaloric $\Delta T$ of $10.2$~K for a tensile $\Delta \sigma$ of $0.55$~GPa \cite{caloric-mapi2}. Although these figures are not directly comparable to our barocaloric $\Delta T_{\rm BC}$ results (i.e., hydrostatic pressure involves isotropic compression whereas tensile uniaxial strain conveys unidirectional stretching), they are of similar size and confirm the promising mechanocaloric potential of MAPI.  

Finally, we note that the barocaloric $\Delta S_{\rm BC}$ results shown in Fig.~\ref{fig6}a differ greatly from the phase entropy changes, $\Delta S_{t}$, estimated with the Clausius-Clapeyron method (Sec.~\ref{subsec:cc}). Firstly, for a pressure shift of $0.1$~GPa, for example, the estimated barocaloric isothermal entropy change is larger than the corresponding phase transition entropy change by about $\approx 10$\%. And secondly, under increasing pressure $\Delta S_{t}$ gets progressively reduced (Table~\ref{table1}) whereas $\Delta S_{\rm BC}$ steadily increases (Fig.~\ref{fig6}a), thus rendering opposite $p$-induced behaviours. Therefore, the Clausius-Clapeyron method, although it may provide an approximate order of magnitude for $\Delta S_{\rm BC}$ at very low pressures \cite{Sau2021,hui22}, is not adequate for estimating BC effects in plastic crystals. (The fact that $\Delta S_{t} \neq \Delta S_{\rm BC}$ is well established in experimental works \cite{lloveras19,li19,aznar20,aznar21}, however, it probably needs to be more emphasized in the context of theoretical calculations.)

\section{Discussion}
\label{subsec:discussion}
An interesting question to answer for MAPI, or any other orientationally disordered crystal, is the following: which are the partial contributions to the order-disorder phase transition entropy change stemming from the vibrational and molecular orientational degrees of freedom? The computational approach introduced in this work can provide a quantitative answer to this question, which is shown in Fig.~\ref{fig7}. At zero pressure, it is found that $\Delta S_{\rm vib}$ is practically twofold larger than $\Delta S_{\rm ori}$, namely, $21.66$ and $11.02$~J~K$^{-1}$~kg$^{-1}$, respectively. (It is noted that the $\Delta S_{t} = \Delta S_{\rm vib} + \Delta S_{\rm ori}$ value obtained in this case is slightly larger than the corresponding entropy change estimated with the Clausius-Clapeyron method --Table~\ref{table1}-- due to the inherent numerical inaccuracies of the latter method explained in Sec.~\ref{subsec:cc}.) Under increasing pressure, the vibrational entropy change gets effectively reduced whereas $\Delta S_{\rm ori}$ remains less affected. In spite of this behaviour, at the highest pressure considered in this study $\Delta S_{\rm vib}$ continues being noticeably larger than the molecular orientational entropy change, namely, $12.96$ versus $8.13$~J~K$^{-1}$~kg$^{-1}$ (Fig.~\ref{fig7}). This result appears to be consistent with recent experimental findings reported for the archetypal plastic crystal adamantane \cite{walker23} and the orientationally disordered ferroelectric ammonium sulfate \cite{walker22}, for which it has been shown that the vibrational contributions to the order-disorder phase entropy change surpass those resulting from molecular reorientational motion.      

The results presented in this article were obtained from $NpT$--MD simulations performed with the MAPI force field developed by Mattoni and co-workers \cite{mattoni_model1,mattoni_model2,mattoni_model3}. To assess the accuracy of this classical interatomic potential in providing correct molecular orientational entropies, we carried out complementary \textit{ab initio} molecular dynamics (AIMD) simulations based on density functional theory \cite{Cazorla2017} (DFT, Methods). Supplementary Fig.~4 shows the angular probability densities obtained for the molecular MA cations at temperatures of $250$ and $400$~K and zero pressure from AIMD simulations. Under the same thermodynamic conditions, the $S_{\rm ori}$ values obtained from $NpT$--MD and AIMD simulations (without offsetting) agree remarkably well, namely, within $1$\%. This finding, on one hand, corroborates the reliability of the molecular orientational entropy results presented above and, on the other hand, suggests that intrinsically accurate first-principles methods, although being computationally very expensive, may be feasibly employed for the analysis of order-disorder phase transitions in orientationally disordered crystals.    

Finally, we comment on the reasonableness of employing analytical free rotor models to describe $S_{\rm ori}$ in orientationally disordered molecular crystals \cite{li20,oliveira23}. From textbooks, it is well known that the entropy of a non-symmetrical molecular free rotor can be analytically expressed as \cite{pathria1972}: 
\begin{equation}
\frac{S_{\rm rot}}{k_{B}} = \frac{3}{2} + \ln{\left[ \frac{8\pi^{2}}{h^{3}} \left( 2 \pi k_{B} T \right)^{3/2} \left( I_{1} I_{2} I _{3} \right)^{1/2} \right]}~,
\label{eq:rotor}
\end{equation}
where $\lbrace I_{i} \rbrace$ are the three principal moments of inertia of the molecule. Since the atomic structure of the methylammonium cation is well established, we can provide here a quantitative and meaningful comparison between the two entropies, $S_{\rm ori}$ and $S_{\rm rot}$, estimated for MAPI. Supplementary Fig.~5 shows the size and temperature dependence of $S_{\rm rot}$ obtained for a molecular MA cation, which are significantly different from the $S_{\rm ori}$ results enclosed in Fig.~\ref{fig4}c. Firstly, $S_{\rm rot}$ monotonically grows under increasing temperature (Eq.~\ref{eq:rotor}) whereas $S_{\rm ori}$ remains practically constant in the disordered phase. Secondly, at room temperature, for instance, $S_{\rm rot}$ is two orders of magnitude larger than $S_{\rm ori}$, namely, $\sim 10^{3}$ and $\sim 10^{1}$~J~K$^{-1}$~kg$^{-1}$, respectively. And thirdly, Eq.~(\ref{eq:rotor}) cannot reproduce any entropy dependence on pressure, even if that is mild, in contrast to $S_{\rm ori}$. The huge quantitative and qualitative differences evidenced here for $S_{\rm rot}$ and $S_{\rm ori}$ can be understood in terms of the facts that in real materials (i)~molecules reorient only along very specific and well defined paths (Fig.~\ref{fig4}b and Supplementary Fig.~3), not along all possible directions like a molecular free rotor does, and (ii)~molecules are not in a perpetual state of orientational motion but instead integrate vibrations and fluctuations around equilibrium states with the actual reorientations. Therefore, as it has been demonstrated in this section, $S_{\rm rot}$ does not seem to conform to a physically well motivated model for understanding and quantifying the orientational entropy and associated changes in plastic crystals and/or hybrid organic-inorganic perovskites \cite{li20,oliveira23}.

\section{Conclusions}
We have introduced a computational approach based on molecular dynamics simulations that allows to investigate barocaloric effects in crystals resulting from molecular order-disorder phase transitions. Both the barocaloric isothermal entropy and adiabatic temperature changes can be quantified from entropy curves expressed as a function of pressure and temperature, just like it is done in quasi-direct barocaloric measurements. The presented simulation method automatically provides the partial contributions to the crystal entropy stemming from the vibrational and molecular orientational degrees of freedom. As a case study, we applied our barocaloric computational approach to MAPI, a technologically relevant perovskite compound that experimentally undergoes an order-disorder phase transition near room temperature at normal pressure. We found giant barocaloric isothermal entropy and adiabatic temperature changes of the order of $\sim 10$~J~K$^{-1}$~kg$^{-1}$ and $\sim 10$~K, respectively, for moderate pressure shifts of $\sim 0.1$~GPa (although these figures probably are somewhat overestimated due to the evidenced limitations of the employed force field). Interestingly, the vibrational degrees of freedom, and in particular those ascribed to the molecular cations, were found to contribute most significantly to the phase transition entropy change at all considered pressures. Furthermore, we demonstrated that the well known analytical model for a molecular free rotor may not be adequate to physically understand and quantify the entropy changes occurring in orientationally disordered materials. We expect that the simulation approach introduced in this study will be broadly adopted by researchers working in the fields of caloric effects and/or disordered materials, thus making a potential great impact in the disciplines of energy materials and condensed matter physics. 
\\

\section*{Methods}
\label{methods-2}
\subsection{Classical molecular dynamics simulations}
\label{subsec:md}
We used the LAMMPS simulation code \cite{lammps} to perform systematic classical molecular dynamics simulations in the $NpT$ ensemble for bulk MAPI by using the force field developed by Mattoni and co-workers \cite{mattoni_model1,mattoni_model2,mattoni_model3}. In these simulations, the temperature was steadily increased up to a targeted value during $1$~ns and subsequently the system was equilibrated during an additional $1$~ns under a targeted fixed pressure. The production runs then lasted for about $100$~ps with $\Delta t = 0.5$~fs, from which the velocities of the atoms and other key quantities (e.g., the potential energy and volume of the system) were gathered. The average value of the temperature and pressure were set by using Nose-Hoover thermostats and barostats with mean fluctuations of $5$~K and $0.05$~GPa, respectively. The simulation box contained $3072$ atoms (equivalent to $256$ MAPI unit cells) and periodic boundary conditions were applied along the three Cartesian directions. The long-range electrostatic interactions were calculated by using a particle-particle particle-mesh solver to compute Ewald sums up to an accuracy of $10^{-4}$~kcal~mol$^{-1}$\AA$^{-1}$ in the forces. The force field uses an hybrid formulation of the Lennard-Jones and Buckingham pairwise interaction models together with long-range Coulomb and harmonic forces (the latter for the molecules). The cutoff distance for the evaluation of the potential energy was set to $10$~\AA. To determine the $p$--$T$ phase diagram of MAPI and its barocaloric performance, we carried out comprehensive $NpT$-MD simulations in the temperature range $180 \le T \le 340$~K, taken at intervals of $10$~K, and in the pressure range $0 \le p \le 0.5$~GPa, taken at intervals of $0.1$~GPa. 
\\

\subsection{\textit{Ab initio} molecular dynamics simulations}
\label{subsec:aimd}
First-principles calculations based on density functional theory (DFT) \cite{Cazorla2017} were performed to analyze the orientational properties of MAPI at zero pressure. The DFT calculations were carried out with the VASP code \cite{kresse1993ab} by following the generalized gradient approximation to the exchange-correlation energy due to Perdew \textit{et al.} (PBE) \cite{Perdew19963865}. The projector augmented-wave (PAW) method was used to represent the ionic cores \cite{Blochl199417953}, and the following electronic states were considered as valence: $2s^22p^2$ for C, $2s^22p^3$ for N, $2s^23p^5$ for I, $1s^1$ for H and $5d^{10}6s^26p^2$ for Pb. We performed \textit{ab initio} MD (AIMD) simulations for large supercells containing $384$ atoms and on which periodic boundary conditions were applied along the three Cartesian directions. The plane-wave basis set was truncated at $500$~eV and for integrals within the first Brillouin zone we employed $\Gamma$-point sampling. The AIMD simulations typically lasted for about $200$~ps and the employed time step was equal to $1.5$~fs. A Nose-Hoover thermostat was used to constrain the temperature in our AIMD calculations in which we mimicked the tetragonal ($I4cm$) and cubic ($Pm\bar{3}m$) phases of MAPI. The dimensions of the system were constrained to the equilibrium volume determined at zero temperature, thus thermal expansion effects were disregarded in our AIMD simulations.
\\

\bibliography{biblio.bib}

\section*{Acknowledgements}
The authors acknowledge TotalEnergies RD High Performance Computing Center in Houston and Production High Performance Computing Center in Pau for the use of our HPC. This work was also supported by the MINECO Projects PID2020-112975GB-I00 and TED2021-130265B-C22 (Spain) and by the DGU Project 2021SGR-00343 (Catalonia). C.C. acknowledges financial support from the Spanish Ministry of Science, Innovation and Universities under the ``Ram\'on y Cajal'' fellowship RYC2018-024947-I. C.E.S. acknowledges financial support from the Spanish Ministry of Science, Innovation and Universities under the "Becas Margarita Salas para la formaci\'on de doctores j\'ovenes" fellowship 2021UPC-MS-67395. Additional computational support was provided by the Red Espa\~nola de Supercomputaci\'on (RES) under the grants FI-2022-1-0006, FI-2022-2-0003 and FI-2022-3-0014. 
\\

\section*{Conflict of Interest}
The authors declare no conflict of interest.
\\

\section*{Data Availability Statement}
The  data  that  support  the  findings  of  this  study  are  available  from  the  corresponding authors upon reasonable request.

\end{document}